\documentclass[letterpaper, 10 pt, conference]{ieeeconf}  

\IEEEoverridecommandlockouts                              
\usepackage{graphicx} 
\usepackage{amssymb}  
\usepackage{theorem}  
\usepackage{psfrag}  
\usepackage{bm}  
\usepackage{algorithmic}
\usepackage{cite}
\usepackage{amsmath}
\usepackage{algorithm}
\usepackage{array}
\usepackage{mdwmath}
\usepackage{mdwtab}
\usepackage{eqparbox}
\usepackage{bm}   
\newtheorem{theorem}{Theorem}[section]

\newtheorem{proposition}[theorem]{Proposition}

\newcommand{\qed}{\nobreak \ifvmode \relax \else
      \ifdim\lastskip<1.5em \hskip-\lastskip
      \hskip1.5em plus0em minus0.5em \fi \nobreak    
      \vrule height0.5em width0.5em depth0.25em\fi}  
      
\newcommand{\ddt}{\frac{\textnormal{d}}{\textnormal{dt}}}

\usepackage{setspace}

\title{\LARGE \bf
Computationally Efficient Trajectory Optimization\\ for Linear Control Systems with Input and State Constraints 
}

\author{Jean-Fran\c cois Stumper and Ralph Kennel, {\it Senior Member}, {\it IEEE}
\thanks{This work was supported through the National Research Funds of Luxembourg under grant PhD-08-070.}
\thanks{J-F. Stumper and R. Kennel are with the Institute of Electrical Drive Systems and Power Electronics, Department of Electrical Engineering and Information Technology, Technische Universit\"at M\"unchen, Arcisstr. 21, D-80333 Munich, Germany. {\tt\small jean-francois.stumper@tum.de}}%
}

\begin{document}

\maketitle
\thispagestyle{empty}
\pagestyle{empty}


\begin{abstract} %
This paper presents a trajectory generation method that optimizes a quadratic cost functional with respect to linear system dynamics and to linear input and state constraints. The method is based on continuous-time flatness-based trajectory generation, and the outputs are parameterized using a polynomial basis. A method to parameterize the constraints is introduced using a result on polynomial nonpositivity. The resulting parameterized problem remains linear-quadratic and can be solved using quadratic programming. The problem can be further simplified to a linear programming problem by linearization around the unconstrained optimum. The method promises to be computationally efficient for constrained systems with a high optimization horizon. As application, a predictive torque controller for a permanent magnet synchronous motor which is based on real-time optimization is presented. 
\end{abstract}

\section{Introduction}


Trajectory optimization in real-time is an important part of many modern control systems, for instance, predictive control. The trajectory optimization problem must be solved in the sampling interval, determined by the plant dynamics. A computationally efficient solution, however, encounters two major obstacles: first, the optimization horizon has a minimum length, either to avoid suboptimality or as stability criterion, and second, the trajectory must satisfy input and state constraints to be feasible. For fast systems, many of the existing constrained predictive control schemes, which are based on numerical iterations, are not applicable. 


Concerning the horizon length problem, the continuous approach to predictive control is of interest \cite{FM00}. Using differential flatness, the optimal control problem can be rewritten as output optimization problem. The basis function approach is applied to obtain a finite-parameter optimization problem \cite{P69} \cite{GF06} \cite{Petit} \cite{NM98}, analogeously to discrete-time optimization. A long optimization horizon is, however, obtained with comparably few optimization parameters.




A remaining issue regarding computational efficiency is the inclusion of constraints. The classical method is the application of penalty functions \cite{GF06}. As an alternative, a coordinate transformation was proposed to modify the constrained problem to a nonlinear unconstrained problem \cite{GP09}, solvable with nonlinear calculus of variations methods. The existing optimization methods with penalty functions or nonlinear coordinate transformations lead to a nonlinear or non-linear-quadratic problem, increasing the computational burden. If only feasibility is of interest, numerical procedures applying time scaling to slow down some variables can be used \cite{FM00}. 



This paper treats the special case of the linearly constrained linear-quadratic problem. The linear structure of the system is exploited in the transformation of the problem to a finite-parameter optimization problem. The linear-quadratic structure of the problem is maintained. The unconstrained problem is solved algebraically as it is convex. A quadratic or linear programming (QP/LP) solver is then applied for constraint handling to modify the solution to a feasible trajectory. The solution is computationally efficient, as first a continuous parameterization is performed, and second a linear programming solver is used for the quadratic problem by linearizing around the unconstrained optimum. To the knowledge of the authors, QP and LP have so far not been applied to this type of continuous problems.

The paper is organized as follows. Section II defines the optimization problem and presents inversion-based linear-quadratic trajectory optimization using basis functions, with the special case of power series. Section III proposes the novel constraint handling method, first using a result on polynomial nonnegativity, second rewriting the problem as linear programming problem. Section IV shows numerical results of a synchronous motor predictive torque controller, which is a multivariable system with a high sampling rate and both input and state constraints. The algorithm is computable in real time: $2$ ms prediction are implemented at a sampling rate of $10$ kHz. The experimental results are presented in \cite{EPE}.


\section{Preliminaries: Continuous Flatness-Based Linear-Quadratic Trajectory Optimization}

\subsection{Optimal Control Problem}

The optimal control problem considered is to generate trajectories on a finite time horizon $T$ minimizing the quadratic cost functional
\begin{align}
J &= \int_{0}^{T} \left( \textbf{x}^\text{T}(t) \textbf{Q} \textbf{x}(t) + \textbf{u}^\text{T}(t) \textbf{R} \textbf{u}(t) \right)\ \textnormal{dt} \notag\\ &\;\;\;\;\;\;\;\;\;\;\;\;\;+ (\textbf{x}(T)-\textbf{x}^*)^\text{T}\textbf{P}(\textbf{x}(T)-\textbf{x}^*)   \label{eq:costfunction}
\end{align}
with states $\textbf{x}(t)\in\mathbb{R}^n$, inputs $\textbf{u}(t)\in\mathbb{R}^m$, weight matrices $\textbf{P}\in\mathbb{R}^{n\times n}$, $\textbf{Q}\in\mathbb{R}^{n\times n}$ and $\textbf{R}\in\mathbb{R}^{m\times m}$, and the desired final state $\textbf{x}^*\in\mathbb{R}^n$. The weight matrices are assumed to be positive definite and symmetric. Any quadratic functional is eligible, for instance the reference $\textbf{x}^*$ can also be included in the cost integral. Quadratic cost functionals are preferred in predictive control as they will provide good closed-loop behavior \cite{Rao}. They can also describe physical costs better than other norms. Furthermore, if the weight matrices are positive definite, the optimization problem is convex and easily solvable: a unique optimum exists and is found by solving first-order necessary conditions. The states and inputs are constrained to the time invariant linear dynamics of the multi-input multi-output system
\begin{align}
\dot{\textbf{x}}(t) &= \textbf{A} \textbf{x}(t) + \textbf{B} \textbf{u}(t)  \label{eq:system} 
\end{align}
with system matrix $\textbf{A}\in\mathbb{R}^{n\times n}$ and input matrix $\textbf{B}\in\mathbb{R}^{n\times m}$. The system is further assumed to be controllable. The optimization is subject to $N_c$ linear input and state constraints
\begin{align}
\textbf{G}_x \textbf{x}(t) + \textbf{G}_u \textbf{u}(t) + \textbf{g}_0 \le 0 ,\;\; \forall t\in[0,T] \  \label{eq:constraints}
\end{align}
with $\textbf{G}_x\in\mathbb{R}^{N_c\times n}$, $\textbf{G}_u\in\mathbb{R}^{N_c\times m}$, $\textbf{g}_0\in\mathbb{R}^{N_c}$, as well as the $n$ initial conditions
\begin{align}
\textbf{x}(0)=\textbf{x}_0 .  \label{eq:initialconditions}
\end{align}
Terminal constraints could be imposed too, but may yield an unfeasible problem in the presence of input or state bounds.

\subsection{Parameterization of the System Variables using Flatness}

By definition, a linear system is said to be differentially flat if there exist $m$ output functions
\begin{align}
\textbf{y}_{f}(t) = \textbf{C}_{f} \textbf{x}(t) ,
\end{align}
with $\textbf{y}_f(t)\in\mathbb{R}^m$, $\textbf{C}_f\in\mathbb{R}^{m\times n}$, such that all states $\textbf{x}(t)$ and inputs $\textbf{u}(t)$ can be expressed as linear combination of the flat outputs and a finite number of their derivatives \cite{FLMR95} \cite{HSR04}. Differential flatness can be interpreted as transformation into controller canonical form, and in the linear case, it is equivalent to controllability \cite{HSR04}. The flat outputs $\textbf{y}_{f}(t)$ are thus the controller canonical form outputs, and the canonical form state vector is
\begin{align}
\textbf{z}(t) &= \left(  \left(  y_{f1}, \dot{y}_{f1}, ..,  y_{f1}^{(r_1-1)} \right),  \left(  y_{f2}, \dot{y}_{f2}, ..,  y_{f2}^{(r_2-1)} \right),  \right. \notag \\
      &\;\;\;\;\;\;\;\;\;\;\;\;\;\;\;\;\;\;\;\;\;\;\;\;\;\;\;\;\;\; \left. ..,    \left(  y_{fm}, \dot{y}_{fm}, ..,  y_{fm}^{(r_m-1)} \right)   \right)^\text{T}     \label{eq:z}
\end{align}
with $\textbf{z}(t)\in\mathbb{R}^n$ and where $r_i$ is the corresponding vector relative degree with regard to the output $y_{fi}$ ($y_{fi}$ being the $i$-th flat output resp. $i$-th element of $\textbf{y}_{f}$ in (5), see \cite{HSR04}). Note that $\sum_{i=1}^{m}r_i=n$. The differential parameterization of the system variables is \cite{FLMR95} \cite{HSR04}
\begin{align}
\textbf{x} &= \bm{\Xi}_x \left( \textbf{y}_f, \dot{\textbf{y}}_f, .., \textbf{y}^{(k-1)}_f \right) ,  \label{eq:flatparamx} \\
\textbf{u} &= \bm{\Xi}_u \left( \textbf{y}_f, \dot{\textbf{y}}_f, .., \textbf{y}^{(k)}_f   \right) ,  \label{eq:flatparamu}
\end{align}
where $k\in\mathbb{R}$ is $k=\max\{r_i\}$. The functions $\bm{\Xi}_x$ and $\bm{\Xi}_u$ are linear functions of $\textbf{z}$ resp. of $\textbf{z}$ and $\dot{\textbf{z}}$, as they are derived from the controller canonical form transformation. See that (\ref{eq:flatparamx}) is the inverse transformation from controller canonical form, and (\ref{eq:flatparamu}) follows from (\ref{eq:flatparamx}) and (\ref{eq:system}).


\subsection{Output Parameterization with a Polynomial Basis}

The Ritz-Galerkin method, also called basis function approach, is a direct method to find an approximate solution to an optimal control problem \cite{KK64}. In the flatness-based approach, the outputs are parameterized as a linear combination of time-variant basis functions. It is comparable to the control parameterization Ritz method (CPRM) \cite{Sirisena73}, where the inputs $\textbf{u}$ are parameterized. Convergence of this method (combined with linear splines) was studied in \cite{Sirisena79} and more generally in \cite{Bosarge}.

As basis functions, polynomials are chosen. Their simplicity is exploited for the further developments, main advantage being that the parameters enter linearly, for instance they allow to prove that the transformed cost remains convex. They also allow constraint transformation based on a result on polynomial nonpositivity in section III. Furthermore, the dynamic shape of the resulting trajectory can be well analyzed.

Higher-order polynomials (for instance Laguerre and Legendre) have been used in many applications, but here, power series, the simplest form of polynomials, are used. The methods and results are all applicable to higher-order polynomials, there, the undetermined coefficients still enter linearly and they might inherit numerical advantages (power series are numerically stable only up to degree $12$..$15$), but the notation would be less comprehensible. In \cite{Brown} it was shown that only the polynomial degree but not the type of series are important for convergence. The choice is a simple approximate, but yet, it turns out to be quite accurate.

The system outputs are defined as degree $N$ power series
\begin{align}
y_{fi}(t) = \sum_{j=0}^{N} \alpha_{ij} \ \frac{t^{j}}{T^j}   \ \ , i=1..m, \ \ t\in[0,T]    \label{eq:trajectory}
\end{align}
with $\alpha_{ij}\in\mathbb{R}$. This is not an approximation of a system step response, but merely an assumption that the optimal solution can be described as a polynomial. The system response is exact, only the output is reduced in dimensionality.

\begin{figure}[!ht]
  \centering
  \includegraphics[width=8.5cm]{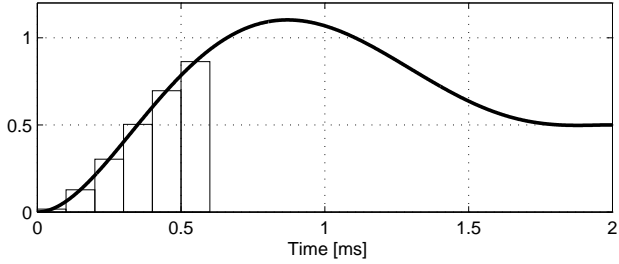}
  \caption{Trajectories with $6$ free parameters. Discrete-time horizon is maximum $0.6$ ms, continuous trajectory is well-conditioned at the desired horizon length of $2$ ms.\label{fig:polynomvseuler}}
\end{figure}

The original problem is thereby transformed to a finite-parameter optimization problem where a set of constants is searched. Fig. \ref{fig:polynomvseuler} demonstrates the computational advantage. For instance, in the example in section IV, the prediction horizon is $2$ ms (the slowest time constant in the system) and the sampling rate is $10$ kHz (determined by the power stage). A discrete description takes $20$ parameters, but here, only $6$ are sufficient to describe a setpoint change. A more complicated trajectory may not be expected on most plants.

The $n$ initial conditions (\ref{eq:initialconditions}) define $n$ of the parameters $\alpha_{ij}$, they are computed via the transformed state $\textbf{z}(0)$ in (\ref{eq:z}) (See that the initial time is $t_0=0$, such that only the first element of the power series is nonzero, same for the derivatives). For further notational simplicity, the vector $\bm\alpha$ is defined as the vector of the undetermined coefficients
\begin{align}
\bm{\alpha} = (\alpha_{1r_1},..,\alpha_{1N},\alpha_{2r_2},..,\alpha_{2N},..,\alpha_{mr_m},..,\alpha_{mN})^\text{T},
\end{align}
with $\bm\alpha\in\mathbb{R}^{N'}$ with $N'=m\times N-n$, and the indices $r_i$ represent the respective vector relative degree of the flat output $y_{fi}$.

Based on the flatness parameterization (\ref{eq:flatparamx}) and (\ref{eq:flatparamu}), and knowing that $\frac{d^i}{dt^i}\textbf{y}_f$ are polynomials, the system variables can be parameterized with the coefficients $\bm\alpha$ yielding the functions
\begin{align}
\textbf{x} (t)  &= \bm{\Gamma}_x(\bm{\alpha},t) ,  \label{eq:diffparamx} \\
\textbf{u} (t)  &= \bm{\Gamma}_u(\bm{\alpha},t) .  \label{eq:diffparamu}
\end{align}
These functions are degree $N$ polynomials of time $t$, where the coefficients $\bm{\alpha}$ enter linearly.

It is seen that $N \ge \max\{r_i\}$, $i=1..m$, is mandatory to have degrees of freedom on the trajectory. It should be remarked that when choosing a higher $N$, the error bounds of the basis function approach become smaller and the dynamical response increases, at the cost of higher computational requirements and the risk of less good numerical conditioning of the parameters.

\subsection{Remark: Generality of the Parameterization for Linear Systems with Power Series Outputs}

It is remarked that the same parameterization (\ref{eq:diffparamx}) and (\ref{eq:diffparamu}) can be obtained for arbitrary non-flat outputs of a controllable and observable linear system, if the output is defined as polynomial series with undetermined coefficients \cite{SK10}. Then, the outputs do not need to be redefined to flat, or controller canonical form outputs. 

\subsection{Parameterizing the Cost Functional}

The parameterization of the system variables (\ref{eq:diffparamx}), (\ref{eq:diffparamu}) is  applied to transform the cost functional (\ref{eq:costfunction}) to a finite-parameter cost function.

\begin{proposition}  \textnormal{\;\\Conditioning the cost functional.} \\
The cost functional (\ref{eq:costfunction}) with the substitutions (\ref{eq:diffparamx}) and (\ref{eq:diffparamu}) yields a biaffine function in the undetermined parameters $\bm\alpha$ of the type
\begin{align}
J(\bm\alpha) =   \bm\alpha^\text{T} \textbf{K} \bm\alpha + \textbf{k}^\text{T} \bm\alpha + k_0    \label{eq:transformedcost}
\end{align}
with $k_0\in\mathbb{R}$, $\textbf{k}\in\mathbb{R}^{N'}$ and $\textbf{K}\in\mathbb{R}^{N'\times N'}$.  \qed
\end{proposition}

The proof follows from a straightforward computation with the knowledge that (\ref{eq:diffparamx}), (\ref{eq:diffparamu}) are polynomials with affine coefficients $\bm{\alpha}$, and is omitted. 

Eq. (\ref{eq:transformedcost}) is derived directly from (1) by substitution of (7), (8) and (9) (or (11) and (12)). Practical (symbolical) calculations are performed using a computer algebra tool, matrix $\textbf{K}$ is the Hessian matrix of $J$, and vector $\textbf{k}$ follows from the gradient of $J$.

\begin{proposition}  \textnormal{\;\\Convexity of the conditioned cost function.} \\
The conditioned cost function (\ref{eq:transformedcost}) is convex over the set $\bm\alpha\in\mathbb{R}^{N'}$ if matrices $\textbf{Q}$, $\textbf{R}$ and $\textbf{P}$ in (\ref{eq:costfunction}) are positive definite and real-symmetric matrices. \qed
\end{proposition}

This statement of convexity is not obvious, as the transformed cost function (\ref{eq:transformedcost}) has a higher dimension as the original cost functional (\ref{eq:costfunction}) ($N'$ instead of $n$). It is proven for the unconstrained linear-quadratic case. This proof follows from the knowledge that the states and inputs are polynomials with coefficients linearly dependent on $\bm\alpha$, and is sketched in appendix I. As only linear constraints are regarded, the result can be assumed valid also for the constrained case.

This result allows to apply first-order conditions of optimality, and will guarantee a solution in finite time as an unique minimum of $J$ exists.

\section{Main Results: Constraint Handling}

The previous section presented the unconstrained optimization of a controllable linear system with a flat output parameterized with a polynomial basis. This section introduces a computationally efficient method for including constraints in trajectory generation. It is a follow-up to the previous section.

\subsection{Parameterizing the Constraints}

The input and state constraints (\ref{eq:constraints}) are transformed with the system variable parameterizations (\ref{eq:diffparamx}) and (\ref{eq:diffparamu}). With the knowledge that the parameterizations (\ref{eq:diffparamx}) and (\ref{eq:diffparamu}) are univariate polynomials in $t$, each of the constraints in (\ref{eq:constraints}) is also a polynomial in $t$ of order $N$ where the coefficients are affine functions in $\bm\alpha$. Thus, the constraints are rewritten as 
\begin{align}
P_k(t)=\sum_{i=0}^{N} \left( g_{ki0} + \textbf{g}_{ki}^T \ \bm\alpha \right) \ t^i \le 0, \;\; k=1..N_c, t\in[0,T] , \label{eq:constraint1}
\end{align}
with $g_{ki0}\in\mathbb{R}$ and $\textbf{g}_{ki}\in\mathbb{R}^{N'}$. Checking constraints over a time interval $t\in[0,T]$ is difficult. The typical solutions for such continuous problems are penalty functions in the cost functional, or nonlinear state transformations. In the following, the constraints are transformed to allow a direct check on $\bm\alpha$ independent of $t$, while maintaining the simplicity of a linear constraint.

\begin{proposition}   \textnormal{\;\\Sufficient affine conditions for the constraints.} \\
The univariate polynomials $P_k(t)$ are positive on the finite interval $t\in[0,T]$ if the conditions
\begin{align}
P_k(0)                           &\le 0,    \;\; \forall  k=1..N_c,  \\
P_k\left(p\frac{T}{N}\right) -\Delta \cdot P_k(0) &\le 0,    \;\; \forall  p=1..N, k=1..N_c,  \label{eq:suffcond}
\end{align}
with a constant $\Delta \in\mathbb{R}>0$ are satisfied.\qed
\end{proposition}

\begin{figure}[!ht]
  \centering
  \includegraphics[height=4.5cm,width=8.5cm]{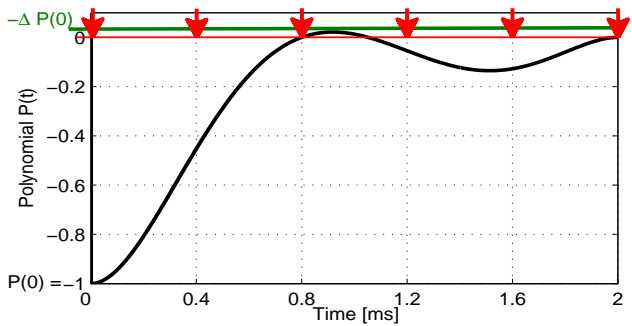}
  \caption{A degree $N$ polynomial trajectory (black) constrained at $N+1$ points to be nonpositive (red arrows) will not exceed the upper bound of $-\Delta \cdot P(0)$ (green line). Here, $N=6$, $P(0)=-1$ and $\Delta=0.037$.\label{fig:nbed}}
\end{figure}

The constraints, written in the introduced notations, are
\begin{align}
\left( g_{k00} + \textbf{g}_{k0}^T \ \bm\alpha \right) \le 0, \;\; k=1..N_c
\end{align}
and
\begin{align}
\sum_{i=0}^{N} \left( g_{ki0}+\textbf{g}_{ki}^T\bm\alpha) \frac{T^i}{N^i} p^i  - \Delta \cdot (g_{k00} + \textbf{g}_{k0}^T \ \bm\alpha \right)  \le 0,\notag\\
    p=1..N, k=1..N_c .
\end{align}
The underlying idea is shown in Fig. \ref{fig:nbed}. If a polynomial trajectory $P(t)$ of degree $N$ is constrained at $N+1$ sampling points on the interval $[0,T]$ to be nonpositive, then, in the worst case, this polynomial reaches a maximum of $P(t)=-\Delta \cdot P(0)$. This maximum offset of $-\Delta \cdot P(0)$ is applied as interleaf to the constraint boundary (the second term in (\ref{eq:suffcond})). The proof of this proposition is presented in appendix II, where also the value of the constant $\Delta$ is found, which is a constant depending on the polynomial degree $N$ only. After transformation of the constraints, they can be used directly in an optimization procedure, as the result is an affine and purely parametric constraint. They are affine functions of $\bm{\alpha}$.

The resulting set of constraints is sufficient, but not necessary, thus they guarantee maintainance of the constraints, but are too restrictive. The offset between the necessary and the sufficient conditions is the factor $\Delta$, and the restriction is, for example, $6.4\%$ for $N=3$, $1.2\%$ for $N=10$, and even less for higher $N$. It is assumed that this restriction is acceptable for most applications regarding the inherited computational advantages.

Exact results on necessary and sufficient conditions on positivity of univariate polynomials over an interval $[0,T]$ exist. Semidefinite programming (SDP) is applied to establish linear matrix inequalities (LMI), which, just as for the presented results, are independent of $t$ and can directly be used in a program \cite{HL03}. These methods, however, require a numerical iterative algorithm to establish the LMIs, and can therefore not be considered if computational efficiency is important. Due to updated initial conditions, the constraints must be reconditioned in each trajectory generation cycle, and the computational burden adds up.



\subsection{Trajectory Generation in Finite Time}

In the previous section, the optimal control problem has been transformed to a finite-parameter optimization problem. The transformed problem can be solved in finite time using quadratic programming (QP) to find the parameters $\bm{\alpha}$, as the cost function is quadratic in $\bm{\alpha}$ and as the constraints are affine functions of $\bm{\alpha}$. This is an interesting result for continuous systems, as QP is a standard method in discrete-time optimization. Furthermore, the number of parameters is decoupled from the horizon length, thus higher prediction horizons can be reached with less parameters. 

An even more convincing advantage of the presented method shall be a further reduction of the computational complexity. In the following, the problem will be approximated such that it is solvable using linear programming (LP) techniques. The approximation yields a feasible solution (thus it is satisfying the constraints) with a bounded suboptimality.

\subsection{Transformation to a Least Distance Problem}

In the first step, a transformation is performed to obtain a simpler cost functional.

\begin{proposition}   \textnormal{\;\\Reformulation as least-distance problem.} \\
The cost functional can be transformed to
\begin{align}
J = \textbf{f}^\text{T}\textbf{f} + c ,
\end{align}
where $\textbf{f}=(f_1,f_2,..,f_{N'})^T\in\mathbb{R}^{N'}$ is the vector of transformed parameters and $c \in\mathbb{R}$ a constant.\qed
\end{proposition}

The new parameter vector $\textbf{\textit{f}}$ is defined as
\begin{align}
\textbf{\textit{f}} = \textbf{F}(\bm\alpha -\bm\alpha_0) , \label{eq:trafoleastdistance}
\end{align}
where $\bm\alpha_0\in\mathbb{R}^{N'}$ is the unconstrained minimum of the parameterized cost functional (\ref{eq:transformedcost})
\begin{align}
\bm\alpha_0 = -\frac{1}{2} \textbf{K}^{-1} \textbf{k} ,
\end{align}
and matrix $\textbf{F}\in\mathbb{R}^{N'\times N'}$ is defined such that
\begin{align}
\textbf{F}^\text{T}\textbf{F} = \textbf{K}
\end{align}
with $\textbf{K}$ from eq. (\ref{eq:transformedcost}). Computation is not an issue as $\textbf{K}$ is positive definite and symmetric, for instance, in the notations in appendix I one would set $\textbf{F} = \textbf{B} \textbf{A}$. If this decomposition is not known, $\textbf{F}$ can be computed via the Cholesky matrix decomposition $\textbf{F}^\text{T}=\mathrm{cholesky}(\textbf{K}^\text{T})$. 

Transformation (\ref{eq:trafoleastdistance}) must also be applied to transform the constraints. This is done by substituting the inverse transformation
\begin{align}
\bm\alpha = \bm\alpha_0 + \textbf{F}^{-1}\textbf{\textit{f}}  \label{eq:retrafoleastdistance}
\end{align}
to the constraint equation, yielding directly the affine constraints in terms of $\textit{\textbf{f}}$. The same equation is applied to retransform the obtained results into the original parameters. Contrary to the amount of parameters, the amount of constraints is not increased.

The least-distance problem is simpler to solve as the original QP problem, as some computations become obsolete in the iterations \cite{F87}. The transformation renders the quadratic programming more efficient and increases numerical stability.


\subsection{Transformation to a Linear Programming Problem}

Now the least distance problem can be approximated for a solution using linear programming. The $L_2$-norm is rewritten as $L_1$-norm. The variables in linear programs are limited to positive numbers, thus the variables $\textbf{\textit{f}}$ are replaced by
\begin{align}
f_i = f_{ip} - f_{in} , 
\end{align}
with $f_{ip}, f_{in} \in \mathbb{R}$  and $f_{ip}\ge0$, $f_{in}\ge0$.

\begin{proposition}   \textnormal{\;\\Linear Cost Function for the Least-Distance Problem.} \\
The cost function approximation
\begin{align}
J = c+ \sum_{i=1}^{N} f_i^2  \approx  c+ \sum_{i=1}^{N} | f_i |  =  c+ \sum_{i=1}^{N} \left( f_{ip} + f_{in} \right) = J', 
\end{align} 
with $f_{ip}, f_{in} \ge 0$, yields a feasible solution with bounded suboptimality for a least-distance problem. \qed
\end{proposition}

This approximation implies a large offset between the linearized cost $J'$ and the correct cost $J$. However, this is not of importance, as the goal is to find a point in the parameter space $\textbf{\textit{f}}$ that is feasible and least-distance to the origin. Under this aspect, the suboptimality is not the variation of the cost, but of the difference of the found parameters in the squared distance to the origin. The contour lines of the quadratic cost are circles around the origin, whereas these of the linear cost are lozenges \cite{Rao}.

It should be remarked that $|f_i|$ is not necessarily equivalent to $f_{ip} + f_{in}$, but in the reverse, if $f_{ip} + f_{in}$ is minimized, it is equivalent to $|f_i|$ as $f_{ip}, f_{in} \ge 0$, such that at least one of each $f_{ip}, f_{in}$ is zero.

The advantages of a linear programming solver are obvious, increased reliability and reduced computational complexity, and the computational burden only grows linearly with the degree of the output polynomial $N$. It represents a significant computational saving compared to when applying QP as solver.

If no constraint is active, the solution is exact. If a constraint is active, the worst case is when the active constraint vertex is parallel to a contour line (i.e. parallel to a side of the lozenge). It can be shown that the resulting cost is
\begin{align}
J' = J_0 + N' \times J_{C}
\end{align}
in the worst case, where $J'$ is the cost with the linear program, $J_0$ the cost of the unconstrained problem, $J_C$ the extra cost when considering constraints in a quadratic program, and $N'=\dim(\bm\alpha)$ the amount of free parameters $\bm{\alpha}$. Thus, the linear programming solver yields the worst-case suboptimality $J'/(J_0+J_C)$ of up to $N'$ times the cost. If not all constraints are active, or if the constraints intersect the contour lines, suboptimality will be considerably less. The bound of the suboptimality and the consideration that the overall suboptimality is relative to the cost of the unconstrained problem makes the simplification of a linear programming solver acceptable for many applications.


\section{Example: Predictive Torque Control of Permanent-Magnet Synchronous Machines}

To present the advantages and the good functionality of the presented trajectory optimization scheme, a predictive torque controller for a  permanent-magnet synchronous machine (PMSM) is presented. So far, this plant has been controlled via generalized predictive control (GPC), via finite control set MPC (FS-MPC) and via the explicit solution \cite{Cortes}. Even though the methods provided good results, they were either unconstrained, or limited to few prediction steps. The presented method obtains a high optimization horizon ($2$ ms) while respecting all constraints. 

Its suboptimality in 1) the assumption of a polynomial output, 2) the restrictive constraint parameterization (of $2\%$), and 3) the use of LP instead of QP is analyzed.

\begin{figure}[!ht]
  \centering
  \includegraphics[height=4cm]{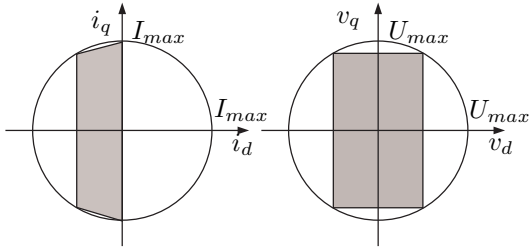}
  \caption{Linearized current and voltage constraints. Circle: feasible set of current and voltage vectors, grey: feasible set after linearization of the constraints.\label{fig:constraints}}
\end{figure}

A cascaded control structure is chosen \cite{Delaleau}. The speed loop is controlled via a PI controller, and the current (and torque) loop is the predictive controller. The electrical subsystem of a non-salient PMSM, consisting of the torque-generating and field-generating currents $i_q$ and $i_d$ (peak values), and the machine torque $\tau_M$ as output, is given as
\begin{align}
\ddt i_d &=  -\frac{R}{L} i_d  +n_p \omega_M i_q  +\frac{1}{L} v_d    \\
\ddt i_q &=  -\frac{R}{L} i_q  -n_p \omega_M i_d  -n_p \omega_M K  +\frac{1}{L} v_q  \\
  \tau_M &=   \frac{3}{2} n_p K i_q
\end{align}
which is linearized by the assumption that speed is constant over the prediction horizon,
\begin{align}
\ddt \omega_M(t_0) &\approx 0 \;\;\; \Rightarrow \omega_M(t)=\omega_M(t_0) \;\;\; \forall t\in[t_0,t_0+T] .
\end{align}
The parameters are taken from a real machine: $R=0.86$ $\Omega$, $L=6$ mH, $n_p=3$, $K=0.236$ Vs, rated speed $314$ rad/s, rated torque $10$ Nm. The trajectory optimization shall minimize the control error
\begin{align}
P_{ctrl}(t) = (\tau_M(t)-\tau_M^*)^2
\end{align}
and at the same time optimize the power efficiency by minimizing the dissipated energy
\begin{align}
P_{loss}(t) &= R \left( i_d^2+i_q^2 \right) +\frac{\omega_M}{R_m} \left( (Li_d+K)^2+i_q^2 \right), \label{eq:Ploss}
\end{align}
where $R_m=1800\Omega$, thus the goal is to minimize $J=\int_0^T (qP_{ctrl}(t)+P_{loss}(t))dt +q\,T\, P_{ctrl}(T)$ with the weight $q=20$. The flat outputs $i_d$ and $i_q$ are each parameterized with a degree $5$ power series. The prediction horizon is $T=2$ ms and the sampling rate is $10$ kHz. The optimization must also maintain the current constraints 
\begin{align}
i_d^2+i_q^2 \le I_{max}^2 = 10^2 \textnormal{A}^2
\end{align}
to protect the power inverter, and the voltage constraints
\begin{align}
v_d^2+v_q^2 \le V_{max}^2 = 330^2 \textnormal{V}^2
\end{align}
to obtain a feasible trajectory. These constraints are linearized as shown in Fig. \ref{fig:constraints}, similar as in \cite{Bolognani}. The argumentation is that only $i_d\le0$ makes sense (field-weakening, see (\ref{eq:Ploss})), and that the current and voltage range in $q$-axis is more important for PMSMs with isotropic rotors.

\psfrag{#iq}{$i_q$}
\psfrag{#id}{$i_d$}
\psfrag{#uq}{$v_q$}
\psfrag{#ud}{$v_d$}
\psfrag{#imx}{$I_{max}$}
\psfrag{#umx}{$U_{max}$}
\psfrag{#-il}{$-\frac{I_{max}}{2}$}

\begin{figure*}[!h]
  \centering
  \includegraphics[width=17cm]{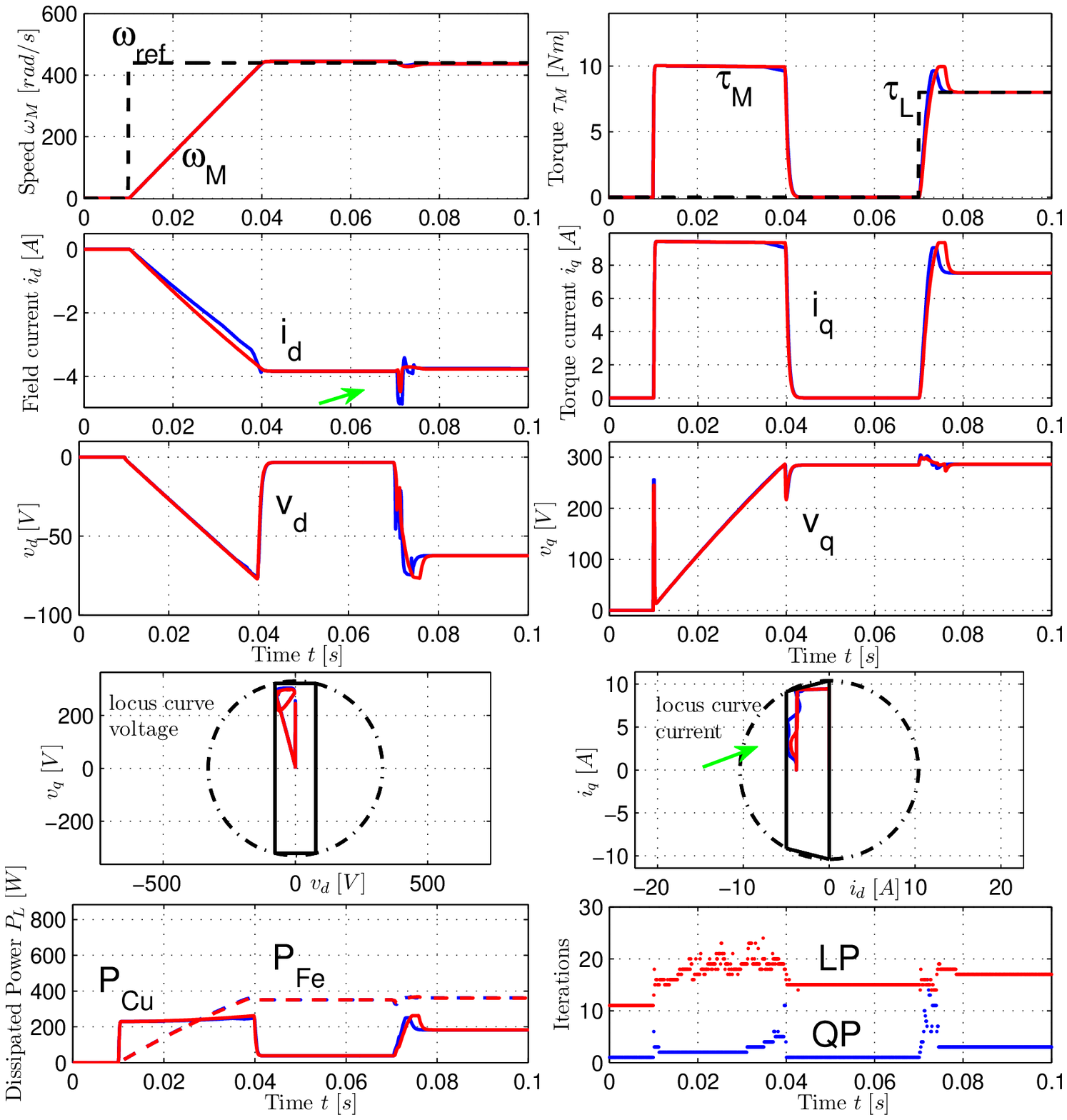} 
  \caption{Results of predictive control applying the presented trajectory optimization. Red: results with LP solver, Blue: results with QP solver.\label{fig:results}}
\end{figure*}

Numerical simulation results are shown in Fig. \ref{fig:results}. The results of the QP solver are shown in blue, and the results of the LP solver in red. Experimental results are shown in \cite{EPE}. At $t=0.01$ s, a speed setpoint change from $\omega_M=0$ to $420$ rad/s is commanded. Then, at $t=0.07$ s, a load torque step of $8$ Nm is applied. From the computational results, QP takes 17 iterations in the worst case, wereas LP takes 24 iterations. This increased number is logical, as there are 20 parameters instead of 10. As the LP algorithm is much simpler, this still represents a considerable computational saving. With an optimized C implementation, the worst case computation time of the LP was $20 \mu$s on a PC ($1.4$ GHz clock). Runtime is further discussed in \cite{EPE}.

From a qualitative standpoint, the resulting behavior is identical for both solvers. The system is always operating at its performance limit. There is a (feasible) voltage peak on $v_q$ at $t = 0.01$ s to rapidly establish a torque current $i_q$, and after this, the induced voltage increases proportional to speed on $v_d$ and $v_q$. The iron losses are reduced by imposing $i_d<0$. Then, at high speed, the load step is quickly compensated via the cascaded speed controller, which commands a torque increase. To do this, the trajectory optimization generates a negative peak on $i_d$ (green arrow) which, according to the model, reduces the induced voltage on $i_q$, thereby $\ddt i_q$ is higher as when keeping $i_d$ small. This is the advantageous behavior of predictive control; in contrast to feedback controllers with saturation, the coupling between the states is exploited to bypass the constraints in an optimal fashion. This result is only visible as a high optimization horizon and the constraints are included in the trajectory generation.

As the constraint handling of LP is suboptimal, there is a quantitative difference, especially in dynamical operation. The peaks are generally somewhat smaller, and at $t=0.07$, the negative $i_d$-peak is of shorter duration and thus less effective.


\section{Conclusion}


A trajectory optimization scheme minimizing a quadratic cost functional to generate continuous trajectories for a linear control system with linear constraints was presented. The scheme recombines several methods in order to obtain a computationally efficient solution. These are the use of differential flatness and of a parameterization using a polynomial basis. A result on polynomial nonnegativity is used as a suitable way to parameterize the constraints. Further developments are done to formulate a linear programming problem. 

Numerical simulations of a predictive controller for a permanent magnet synchronous machine show that the suboptimality of the method is acceptable. The optimization problem is sufficiently simplified and solvable in real-time, in the presented application, $2$ ms prediction are implemented at a sampling rate of $10$ kHz.



\appendices

\small

\section{Proof of Proposition \textbf{III.2}}
\textit{Convexity of the conditioned cost function.} \\
Assume $\textbf{Q}$ is positive definite. Then,
\begin{align}
\textbf{x}^T(t) \textbf{Q} \textbf{x}(t) &> 0 , \;\;\forall \textbf{x}(t) \neq 0  
\end{align}
and
\begin{align}
J = \int_0^T \textbf{x}^T(t) \textbf{Q} \textbf{x}(t) dt &> 0 , \;\;\textbf{x}(t) \neq 0 \forall t \in[0,T]  .
\end{align}
The inverse model replaces $\textbf{x}(t)$ by polynomials $\textbf{p}(t)$ with linear coefficients in $\bm\alpha$
\begin{align}
J = \int_0^T \textbf{p}^T(t) \textbf{Q} \textbf{p}(t) dt &> 0 , \;\; \textbf{x}(t) = \textbf{p}(t) \neq 0 \forall t \in[0,T] .
\end{align}
Define the primitives $\textbf{P}(t)=\int \textbf{p}(t) dt$ thus $\textbf{P}(T)=\int_0^T \textbf{p}(t) dt$
\begin{align}
J = \textbf{P}^T(T) \textbf{Q} \textbf{P}(T) &> 0 , \;\;\forall \textbf{P}(T) \neq 0  .
\end{align}
The expression $\textbf{P}(t)$ is the primitive of a polynomial, thus a polynomial, with linear coefficients in $\bm\alpha$, and $\textbf{P}(T)$ is rewritten as 
\begin{align}
\textbf{P}(T) = \textbf{A} \bm\alpha
\end{align}
and we assume $\mathrm{rank}(\textbf{A})=n$ with $n=\mathrm{dim}(\textbf{x})$. It follows
\begin{align}
J = \bm\alpha^T \textbf{A}^T \textbf{Q} \textbf{A} \bm\alpha &> 0 , \;\;\forall \textbf{P}(T) \neq 0  .
\end{align}
The matrix of the parameterized cost function is $\textbf{K}=\textbf{A}^T \textbf{Q} \textbf{A}$ and as we assumed $\textbf{Q}$ is positive definite we know $\textbf{Q}=\textbf{B}^T\textbf{B}$ (Cholesky decomposition). The weight matrix is then
\begin{align}
\textbf{K}=\textbf{A}^T \textbf{B}^T \textbf{B} \textbf{A} = (\textbf{B}\textbf{A})^T (\textbf{B}\textbf{A})
\end{align}
which is positive definite as any matrix $\textbf{K}=\textbf{C}^T\textbf{C}$ for some $\textbf{C}$ with $\mathrm{rank}(\textbf{C})=n$ is positive semidefinite [C.D. Meyer, Matrix analysis and applied linear algebra, SIAM books, 2000, pp. 566]. The parameterized cost functional $J$ is thus a convex function of the parameters $\bm\alpha$. \qed

\section{Proof of Proposition \textbf{III.3}}
\textit{Sufficient affine conditions for the constraints.} \\
The polynomial
\begin{align}
P(s) = \sum_{i=0}^{N} c_i s^i \le 0  ,  
\end{align}
with $c_i\in\mathbb{R}$, is analyzed on non-positivity over a segment $s\in[0,1]$. A first necessary and sufficient condition is
\begin{align}
P(0) = c_0 \le 0                      \label{eq:neccondproof}
\end{align}
which in the following is assumed satisfied. Furthermore, the $N$ conditions 
\begin{align}
P\left(\frac{k}{N}\right)  \le 0  , \;\; k=1..N  \label{eq:suffcondproof}
\end{align}
are also assumed to hold for all $c_i$.

These conditions can be rewritten in matrix notation
\begin{align}
 \textbf{c}_0 + \textbf{Q} \textbf{c} \le 0 
\end{align}
with $\textbf{c}_0=(c_0,..,c_0)^T \in \mathbb{R}^N$, $\textbf{c}=(c_1,..,c_N)^T \in \mathbb{R}^N$ and $\textbf{Q}\in \mathbb{R}^{N\times N}$ such that
\begin{align}
\textbf{Q} = (q_{ij}) = \left( \frac{\partial}{\partial c_j} P(i/N) \right) = \left(\left(\frac{i}{N}\right) ^j\right), \notag\\
       \;\;\;\;\;\;\;\;\;\;\;\; i=1..N, j=1..N.
\end{align}
It can be shown that $\det(\textbf{Q})\neq 0$ for $N>0$, and that $\textbf{Q}$ is positive definite. It follows 
\begin{align}
 \textbf{c} \le - \textbf{Q}^{-1} \textbf{c}_0 
\end{align}
which can be placed into the polynomial equation
\begin{align}
P(s) = c_0+\textbf{s}^T \textbf{c} \le c_0 - \textbf{s}^T \textbf{Q}^{-1} \textbf{c}_0 = (-c_0) \epsilon  \label{eq:inequality}
\end{align}
with $\textbf{s}=(s,..,s^N)^T \in \mathbb{R}^N$ and $\epsilon=-1 + \textbf{s}^T \textbf{Q}^{-1} (1,..,1)^T$. As we assumed $c_0\le 0$, the upper bound of $P(s)$ under the mentioned conditions is at when $\epsilon$ is at its maximum. It can be shown that the upper bound of $\epsilon$, $\Delta=\sup\{\epsilon\}$ $\forall s\in[0,1]$, is positive and only dependent on $N$, as $\textbf{Q}$ is known. Some values, which were computed numerically, are shown in the table below.\\

\begin{tabular}{|l||c|c|c|c|c|}
\hline
$N$                         &  2     & 3     & 4     & 10    & 20 \\
\hline
$\Delta=\sup\{\epsilon\}$   &  0.125 & 0.064 & 0.041 & 0.012 & 0.005 \\
\hline
\end{tabular}
\;\\

Therefore if the conditions (\ref{eq:suffcondproof}) hold, we have 
\begin{align}
 P(s) \le - \Delta P(0).
\end{align}
Shifting the conditions by the constant (and negative) factor $\Delta P(0)$, the sufficient conditions for non-positivity of the polynomial $P(s)$ are found:
\begin{align}
 P(0) &\le 0  ,  \\
 P(\frac{k}{N}) - \Delta \cdot P(0) &\le 0, \;\;k=1..N.
\end{align}
 \qed

\end{document}